\magnification 1200
\nopagenumbers 
\baselineskip=7mm
\hsize=5.0in

\centerline {\bf {Notes on Choptuik Phenomena}}
\vskip .7truein
\centerline { M.Horta\c csu * }
\vskip .4truein
\centerline {  Physics Department, Faculty of Sciences and Letters}
\vskip .1truein
\centerline { ITU 80626 Maslak, Istanbul, Turkey }
\vskip .1truein
\centerline {and}
\vskip .1truein
\centerline { Feza G\" ursey Institute, \c Cengelk\" oy, 81220, Istanbul, Turkey}
\vskip 1.3truein
\centerline {\bf { Abstract}}
We review the work going on in black-hole physics during the last ten years,  called {\it{ the Choptuik Phenomenon}}.

\vskip 1.5truein
\noindent
* hortacsu@itu.edu.tr
\baselineskip=18pt
\footline={\centerline {\folio}}
\pageno=1
\vfill\eject
\noindent
INTRODUCTION
\vskip.05truein
In these notes I  used mainly the reviews $^{/1-3}$ written by Carsten Gundlach, I freely quoted in this paper sometimes even without quotation marks. If quotation makrks exist without references, these parts are from Gundlach reviews. This gentleman contributed to this field both with his original work and by the three reviews he wrote on this subject.

I start by quoting Gundlach, from the first paragraph of still another paper that can be found in gr-qc/9604019 $^{/4}$.

"In an astrophysical context, gravitational collapse normally starts from a star.  This means that the initial data are almost stationary, and that they have a characteristic scale which is provided by the matter.  Therefore, astrophysical black holes have a minimum mass, namely the Chandrasekhar  mass.  Abandoning the restriction to almost stationary data, or alternatively to realistic matter, one should be able to make arbitrarily small black holes. One may then ask what happens if one tries to make a black hole of infinitesimal mass by fine-tuning the initial data."

We thus learn that for a stationary solution, a finite mass is necessary, but there may be surprises for the non-stationary solutions.
Now I continue quoting Gundlach, but this time from the "Historical Introduction" part of his  second review $^{/2}$.

"In 1987 Christodolou, who was studying the spherically symmetric Einstein-scalar model analytically, suggested to Matt Choptuik, who was investigating the same system numerically, the following question:
Consider a generic smooth one-parameter family of asymptotically flat smooth initial data, such that for large values of the parameter $p$ a black hole is formed, and no black hole is formed for small $p$. If one makes a bisection search for the critical value $p_{*}$ where a black hole is formed, does the black hole have a finite or infinitesimal mass?" "The idea was to start with a matter model that does not admit stable stationary solutions."

"After developing advanced numerical methods for this purpose, Choptuik managed to give highly convincing numerical evidence that the mass is infinitesimal". Moreover he found two totally unexpected phenomena $^{/5}$:

\noindent
The first is the now famous scaling relation
$$ M = C(p- p_{*})^{\gamma} \eqno{1} $$
for the black hole mass $M$ in the limit $p$ close to $ p_{*}$ but greater than $ p_{*}$. Choptuik found $\gamma$ approximately equal to $0.374$ for all 1-parameter families of scalar field data. 

\noindent
The second is the appearance of a highly complicated, scale-periodic solution for $p$ close to $ p_{*}$. If we elaborate on this solution what is happening is the following. "For marginal data, both supercritical and sub-critical, the time evolution approaches a certain universal solution which is the same for all the one-parameter families of data. This solution is an "intermediate attractor" in the sense that the time evolution first converges onto it, but then diverges from it eventually, to either form a black hole or to disperse. This universal solution, also called the "critical solution", has a curious symmetry." The critical solution $\phi_{*}(r,t)$ is the same when we rescale space and time by a factor $e^{\Delta}$:
$$\phi_{*}=\phi_{*}(e^{\Delta}r,e^{\Delta}t). \eqno{2} $$ 
The logarithmic scale period of this solution, $\Delta$ approximately equal to $3.44$ is a second dimensionless number coming out of the blue. 

As a third remarkable phenomenon, both the "critical exponent" and "critical solution" are "universal", that is the same for all one-parameter families ever investigated independent of the shape of the initial data, whether it is a gaussian, a hyperbolic tangent or the two other forms studied in reference  5. For a finite time in a finite region of space, the space-time generated by all near-critical data approaches one and the same solution. This universal phase ends when the evolution decides between the black hole formation and dispersion. The universal critical solution is approached by any initial data that are sufficiently close to the black hole threshold, on either side, and from any 1-parameter family.

Similar phenomenon to Choptuik's results were quickly found in other systems too, even in gravitational waves ( which can form black holes in the absence of matter) suggesting that they were limited neither to scalar field matter nor to spherical symmetry.

The cylindyrically symmetric case was studied by A.M. Abrahams and C.R. Evans $^{/6}$
 with similar results,  having $\Delta$ approximately equal $0.6$ and $\gamma$ approximately equal to $0.36$. For a perfect fluid with equation of state $p= {{\rho}\over{3}}$ $^{/7}$
, the "critical solution" is not discretely, but continuously self-similar (CSS) in spherical symmetry, since the universal attractor has a different symmetry.  $\gamma$ is found to be $0.36$ once more. Other models were also studied with different values for $\gamma$. For models in which fermions are coupled to gravity, similar phenomenon is found $^{/8}$.  From these examples it was clear that $\gamma$ was not universal with respect to different kinds of matter, but only work with respect to the initial data for any one matter model.

When similar systems were studied in lower dimensions, namely in $2+1$, analytical results were also possible. When the system was studied numerically by Choptuik and Pretorius  $^{/9-13}$,
 one obtained 
what seemed to be irrational indices. When Birmingham $^{/14}$
studied the "properties of a massive scalar field in the background geometry of the 2+1 dimensional  black hole", he could solve the wave equation exactly in terms of hypergeometric functions . The remarkable feature was that he could evaluate one of the Choptuik scaling parameters as  the sum and difference of the outer and the inner horizons and show that it was given by a rational number, namely ${{1}\over{2}}$ . This reminds us  that the 2+1 dimensional black hole, so called BTZ black hole, exists in a space which is 'related' to the anti-de Sitter space.  The latter   is related to a conformal model $^{/15-17}$ in one less dimension by the Maldacena correspondence $^{/18}$ 
. The emergence of hypergeometric functions and rational exponents are signatures of "conformal invariance"$^{/19,20}$
. If we consider only the relation of hypergeometric functions to conformal symmetry also found in $^{/21-23}$.
  We will comment on lower dimensional solutions later. 

If we go back to $3+1$ dimensions ,  we see that  interesting phenomenae occur in self-similar solutions.  General related phenomena divide into two categories, the continuosly self-similar solutions like those found by Evans and Coleman,$^{/7}$ where the metric coefficients are of the form
$$g_{\mu \nu} (\tau,x^{i}) = e^{ -2\tau} g*_{\mu \nu}(x^{i}), \eqno {3} $$
and discretely self -similar solutions, found by Choptuik,$^{/5}$
 to where the metric coefficients are of the form
$$g_{\mu \nu} (\tau,x^{i}) = e^{ -2\tau} g*_{\mu \nu}(\tau,x^{i}). \eqno {4} $$  
The right hand side of this equation is given as
$$ g*_{\mu \nu}(\tau,x^{i})=g*_{\mu \nu} (\tau+\Delta, x^{i}). \eqno {5} $$ 
  Here $\tau$ is the negative of the logarithm of a  space-time scale, $\tau=-log (r^2+t^2) $, and $x_{i}$ are dimensionless.  

Work in related fields still continue. A last paper I found by Choptuik et al is given in gr-qc/0305003 $^{/24}$. 
We must first study the first work on this subject.
\vskip.1truein
\noindent
{\bf{THE SPHERICALLY SYMMETRIC CASE WITH SCALAR FIELD}}
\vskip.05truein
Choptuik $^{/5}$
 studied a massless scalar field $\phi $ minimally coupled to the gravitational metric. 
$$ G_{\mu \nu} = 8\pi T_{\mu \nu} \eqno { 6} $$ 
where LHS side has the Einstein tensor and the RHS has the stress-energy tensor of the scalar field. The scalar field obeys the Klein-Gordon equation written in this background. 

He chose the metric as 
$$ ds^2 =-\alpha^2 (r,t) dt^2 + a^2(r,t) dr^2 +r^2 d\Omega^2 , \eqno { 7} $$
where $d\Omega^2 = d\theta^2 + sin^2 \theta d\phi^2 $ is the metric on the unit 2- sphere. He chose $t$ orthogonal to $r$ to have no $drdt$ cross term in the line element. To fix the coordinate completely, he chose $\alpha=1$ at $r=0$, so that $t$ is the proper time of the central observer at $r=0$.

He used auxiliary variables
$$ \Phi= \phi_{,r} ,\eqno{8a} $$  $$ \Pi = {{a}\over{\alpha}} \phi_{,t} \eqno { 8b} $$
to make the wave equation a first order system:
$$ \Phi_{,t} = \left( {{\alpha}\over{a}} \Pi\right)_{,r} , \eqno{9 } $$
$$\Pi_{,t} = {{1}\over{r^2}} \left( r^2 {{\alpha}\over{a}} \Phi \right)_{,r}. \eqno{ 10} $$

In spherical symmetry there are four algebraically independent components of the Einstein equations. One of them  can be written in terms of derivatives of the others. The others read
$$ {{a_{,r}}\over{a}}+{{a^2-1}\over{2r}} -2\pi r(\Pi^2 + \Phi^2) =0 , \eqno { 11} $$
$${{\alpha_{,r}}\over{\alpha}} - {{a_{,r}}\over {a}} -{{a^2-1}\over{r}} =0, \eqno {12 } $$
$${{a_{,t}}\over{\alpha}} -4\pi r \Phi \Pi =0 . \eqno { 13} $$
Choptuik  chose to use the first two equations with no time derivatives on the metric parameters. The scalar field evolved forward in time, while $a$ and $\alpha$ were calculated from this result at each new time step, by explicit integration over $r$ starting at $r=0$. The third equation obeyed automatically. He took $a=1$ at $r=0$ to make the spacetime regular at that point. These equations do not have an intrinsic scale, and scale transformation on $t$ and $r$ transform one solution into another. 

There is no gravitational radiation for the spherically symmetric case. The degrees of freedom of the gravitational field depends on the matter fields. Thus the free data for the system will be given in terms of two functions $\Pi(0,r)$ and $\Phi(0,r)$; the presence of gravity does not change this fact. 

Choptuik studied 1-parameter families of such data by evolving the data for many values of the parameter $p$, which he chose to be the amplitude of the Gaussian he took for $\Pi(r,0)$. He could have taken the width or the center of the Gaussian as well with the same result. He took $\Phi(0,r)=0$.

Christodolou $^{/25,26}$
 has shown that for spherically symmetric systems, data sufficiently weak in a well-defined way evolve into a Minkowski-like spacetime, and that strong data forms a black hole $^{/27}$
 . What Choptuik was looking at what happens in between, where the conditions of neither theorem apply.

He found that he could make arbitrarily small black holes by fine-tuning the parameter $p$ ever closer to the black hole threshold. One point was important. Nothing singled out the black hole threshold in the initial data. One could not tell whether one set of data would form a black hole whereas another set infinitesimally close to this one would not. One only had to let the scalar field evolve in time for a sufficiently long time. At the end he showed that one could form black holes as small as $10^{-6}$ times the ADM mass of the spacetime. The power law scaling was obeyed over six orders of magnitude of masses. He found that the scalar wave contracts until curvature values of order $10^{12}$ are reached in a spacetime region of size $10^{-6}$ before it starts to disperse. He found that all near-critical spacetimes, for all families of initial data, looked the same in the intermediate region. They approximate one universal spacetime, called the critical solution 
$$ Z_{*}(kr,k(t-t_{*})).\eqno{14}$$
The accumulation point $t_{*}$ and the factor $k$ depend on the family, but $Z_{*}$ does not. This universal solution has the property that 
$$Z_{*}(r,t)=Z_{*} (e^{n\Delta}r, e^{n\Delta} t) \eqno{15}$$ 
for all integer $n$ and for $\Delta= 3.44 ..$. Here $Z$ stands for the metric coefficients and the scalar field.

All critical points that have been found in black hole thresholds so far have an additional spacetime symmetry . They are either time-independent or scale invariant. These two kinds exhibit different phenomena, the former giving rise to continuous and the latter to discrete self similar solutions.

Those that are time independent have a mass gap. The critical solution is by definition an attractor of co-dimension one, which means that it has precisely one unstable perturbation, with $\lambda_{0}$ greater than zero. All other perturbations decay, having the real part of their eigenvalues $\lambda_{i}$ less than zero. Intuitively, these critical solutions can be thought of as metastable stars , and they typically occur when the field equations set a mass scale in the problem. 

The second type critical solutions occur when there is no scale in the field equations, or when this scale is not dynamically relevant. Many systems, such as a massive scalar field, show both types of critical phenomena in different regions of the space of initial data.

For the scale-invariant system too, this symmetry comes in a continuous or a discrete version. A continuously self-similar (CSS) solution is invariant under an infinitesimal rescaling of both space and time.  A discretely self-similar (DSS) solution is invariant only under rescaling by a particular finite factor, or its integer powers. They are independent of a suitable scale coordinate. They give rise to power-law scaling of the black-hole mass at the threshold.

Similar phenomena occur in statistical mechanics where self-similar solutions near a second order phase transition are important.   At the critical point between a gas and a liquid, and otherwise clear liquid becomes opaque, due to density fluctuations appearing on all scales up to scales much larger than the underlying atomic scale, and including the wavelength of light.  The density discontinuity vanishes as a non-integer power:
$$ \rho_{liquid} - \rho_{gas} =k (T_{*}-T)^{\gamma} . \eqno {16 } $$
Similar behaviour is seen in a ferromagnet where the spontaneous magnetization  goes as  
$$|{\bf{m}}|=c(T_{*}-T)^{\gamma} . \eqno {17  } $$
In both of these expressions $T_{*}$ is the critical temperature where the respective "order parameter" vanishes.   At this point all the quantities that are dimensional in the model are 'irrelevant', and the system is approximately scale-invariant.

Further work was done on the interpretation of the initial Choptuik solution. A most recent study of the global structure of this solution was done by Martin-Garcia and Gundlach $^{/28}$
where "this solution, formed in the gravitational collapse of a scalar field resulting in a naked singularity, was found to be regular at the center to the past of the singularity, and regular at the past lightcone of the singularity. At the future lightcone of the singularity, which is also a Cauchy horizon, the curvature is finite and continuous but not differentiable. To the future of the Cauchy horizon the solution is not unique, but depends on a free function ( the null data coming out of the naked singularity). There is a unique continuation with a regular center  (which is self-similar). All other self-similar continuations have a central timelike singularity with a negative mass $^{/28}$." 
\vskip.1truein
\noindent
{\bf{ SOLUTIONS IN 2+1 DIMENSIONS}}
\vskip.05truein
 In $2+1$ dimensions, where we do not have to stick only to numerical calculations, analytical solutions may be easier to obtain and tell us more about the  underlying mechanism of this phenomenon  . 
General relativity in $2+1$ dimensions, however,  is quite different from that in $3+1$ dimensions, because the degree of freedom of the graviton is given by $d-3$ in any number of dimensions.  This count does not allow any local degrees of freedom to the graviton in $2+1$ dimensions.  One could study models only with interesting global properties if we study pure gravity.  Furthermore, in $2+1$ dimensions the Riemann tensor does not have any other components as those given by the Ricci tensor, i.e. the Weyl tensor which gives half of the components of the Riemann tensor in $3+1$ dimensions  is identically zero. Another tensor, called the Cotton tensor, which exists in $2+1$ dimensions is also identically zero when we use the Einstein-Hilbert action for gravity.  This fact makes it impossible to have a Ricci flat space with non-trivial Riemann tensor components. Due to this fact, there are no gravitational waves. Blackholes can exist only in anti de Sitter spaces $^{/29}$
, rather than asymptotically flat spacetimes. This solution is called the BTZ solution. 

Quoting from Gundlach $^{/1}$ "The negative cosmological constant of the AdS space locally introduces a mass scale into the field equations. It also changes the the global structure of spacetime. Null infinity becomes  a timelike surface. A null ray can go out from the center to null infinity and return while a finite proper time passes at the center. The only consistent boundary conditions for the massless wave equations are the Dirichlet boundary conditions. This means that all outgoing scalar waves are reflected back to the center in a finite time." 
In a later article $^{/9}$
 Choptuik et al. studied "Gravitational collapse in $2+1$ dimensional AdS spacetime". They find what looks like a CSS solution, of the second type ( with mass vanishing at the critical point). "In this respect, one should note one thing. Since the cosmological constant introduces reflecting (Dirichlet) boundary conditions for the scalar field, all of this matter must eventually fall into the black hole. In this sense, the black hole mass is simply the asymptotic mass of the spacetime, and there is no black hole threshold. In numerical calculations, however, one uses initial data with compact support. One can define the black hole threshold as the formation of an apparent horizon before the scalar waves are reflected for the first time. An apparent horizon is the surface where the gradient of $r$ is null. If the cosmological constant is negligible and the mass $M$ is much less than unity, the dynamics is approximately scale invariant and second type solutions may be found." $^{/1}$

Pretorius and Choptuik $^{/9}$ found $ M $ going as $ (p-p_{*})^{2\gamma} $ where $\gamma=1.20$ approximately. Husain and Olivier $^{/10}$ studied the similar system and found $\gamma=.81$. Garfinkle $^{/13}$
studied this case  analytically by setting the cosmological constant to zero, and found an index close to that of Pretorius and Choptuik. Garfinkle and Gundlach $^{/30}$
calculated the perturbation spectrum of Garfinkle's solutions in closed form with the assumption that the perturbations are analytic. They found instead of one growing mode, three growing modes. The dominant one gives the index  for an unstable mode to be equal to ${{8}\over{7}}$ which is in good agreement with the Pretorius-Choptuik work. All this work has uncertainties , though.  Another index found in reference 30 is ${{4}\over {3}}$.  Hirschmann et al $^{/31}$ use a constraint to suppress some of these modes and find the index $\gamma$ equal to four.

Birmingham and Sen $^{/32}$
considered the formation of a black hole from the collision of two point particles of equal mass in $2+1$ dimensions with a negative cosmological constant, which would have the metric given below in the absence of the scalar particles.
$$ ds^2 = -\left(-M+{{r^2}\over{(-\Lambda)}}\right) dt^2+ \left(-M+{{r^2}\over{(-\Lambda)}}\right)^{-1} dr^2 +r^2 d\theta^2 . \eqno{18 } $$
Here M takes values from $-1$ to infinity. $M=-1$ is the anti-de Sitter space. $-1<M<0$ results in point particle naked singularities at the center. When $M$ equals or exceeds zero, we have black holes. 

Birmingham and Sen found the index to be ${{1}\over{2}}$ for the case when the phase space is two dimensional. Peleg and Steif $^{/33}$
studied the collapse of a dust ring where the space of initial data is two dimensional and found the same value for the index. In the Birmingham $^{/34}$
construction the presence of two Virasoro algebras, so the relation to conformal symmetry in two dimensions is evident. He goes further and expands a massive scalar field in the background metric of a rotating black hole:
$$ ds^2 = -\left(-M+{{r^2}\over{(-\Lambda)}}+{{J^2}\over{4r^2}} \right) dt^2+ \left(-M+{{r^2}\over{(-\Lambda)}}+{{J^2}\over{4r^2}}\right)^{-1} dr^2 +r^2 \left(d\theta-{{Jdt}\over{2r^2}}\right)^2 . \eqno{19 } $$
In this background the scalar wave equation can be solved by an ansatz of the form
$$\Phi=R (r) e^{-i\omega t} e^{im\phi} .\eqno{20 } $$
After a change of variables to  $ z={{r^2-r_{+}^2}\over{r^2-r_{-}^2}} $, where $r_{+}$ and $r_{-}$ denote the outer and inner horizon radii, the points when $ (-M+{{r^2}\over{(-\Lambda)}}+{{J^2}\over{4r^2}}) =0$, the radial equation is of the hypergeometric type, with solution $ R(z) = z^{\alpha} (1-z)^{\beta}  _{2}F_{1}(a,b,c,z) $.
The implementation of the vanishing boundary condition at infinity, fixes the constants $a,b,c,\alpha , \beta$ in terms of the parameters of the metric, which can be used to find the index for the mass, after a detour .

As stated above, the presence of hypergeometric functions among the solutions reminds us of the other places where they occur, always related to the presence of conformal symmetry in the problem.  We should note that we get rational indices when we perturb around the exact solution which is related  to a model with conformal invariance $^{15-17}$.

In still another paper Birmingham et al. $^{/35}$
 show how the $2+1$ black hole gives rise to:
\vskip.05truein
\noindent
a)no hair theorem: The Euclidean BTZ black hole has to topology of a torus which is parametrized only by  two real  parameters, which makes one conclude that at most two parameters, mass and angular momentum are sufficient to describe the system. 
At this point also note how one can not construct a charged rotating black hole directly, but have to use an indirect method $^{/36}$.
\vskip.05truein
\noindent
b)holography: BTZ black hole is a holographic manifold, since the three dimensional hyperbolic structure is in 1-1 correspondence with the parameters of the two-dimensional genus one boundary.
\vskip.05truein
\noindent
c)entropy: The Bekenstein-Hawking entropy is determined by the Teichm\" uller space on the boundary.
\vskip.05truein
\noindent 
d)Maldacena Conjecture $^{/18}$:
The three dimensional hyperbolic space is related to a $ SL(2,Z)$ family of solutions.
\vskip.08truein

We will continue this review with a summary of some work $^{/37}$ 
 in $2+1$ dimensions where some of the above mentioned phenomena is displayed a little more clearly. We will show, how in this overly simplified case, the naked singularity is avoided.
\vskip.1truein
\noindent
{\bf{TOY MODEL}} 
\vskip.05truein 
We start with the equations given by Pretorius and Choptuik $^{/9}$
for a metric given by 
$$ ds^2= {{e^{2A(r)}}\over {cos^2 (r) }} (dr^2 - dt^2) + tan^2 (r) e^{2B(r)} d\theta^2 .\eqno {21}$$ 
Here the full space is mapped into the interval $ 0<r < {{\pi}\over{2}} $. The cosmological constant is chosen equal to $-1$ and $r$ is scaled so that it is a dimensionless parameter.  We take the static and the spherically symmetric case, where $A,B$ are functions of $r$ only.  Since we are treating the spherically symmetrical case, $\theta$ independence is justified.   Here, as in all static cases, we expect to have a finite mass.  We are using this example only to show how the singularity is formed, and how it is avoided.

If a scalar, static particle is coupled to the static metric, the Einstein equations, with the cosmological constant read  $^{/9}$
 $$ A_{,rr} + {{(1-e^{2A})} \over { cos^2 (r)}} + 2\pi \phi_{,r}^2 =0 , \eqno {22} $$
 $$B _{,rr}+ B_{,r} \left( B_{,r} +{4 \over {sin(2r)}}\right) + {2{(1-e^{2A})} \over {cos^2 (r)}} =0 , \eqno {23 } $$ 
$$ B_{,rr}+B_{,r} \left( B_{,r}-A_{,r}+2{{(1+ cos^2 (r))} \over {sin (2r) }}\right)- {2A_{,r} \over { sin (2r)}} + {{(1-e^{2A}) } \over {cos^2(r)}} + 2\pi \phi_{,r}^2 =0,\eqno {24} $$
 $$ [ tan r e^B \phi_{,r} ]_{,r} =0 .\eqno {25} $$ 
This system of equations have the set of solutions 
$$ \phi=0, A= -log (sin r), B= log({{cos 2r} \over {2sin^2 r}}) .\eqno {26}$$ 
The expressions for $A$ and $\phi$ were given by Pretorius and Choptuik . We see that our original domain, $0<r< {{\pi}\over{2}}$   is halved.  Due to the singularity in $B$, we can use only the region where $0<r<{{\pi}\over{4}}$. 

We perform a simple perturbation expansion at this point treating the above set of solutions as the zeroth order term. $$ A= A_0+\epsilon A_1 + \epsilon^2 A_2 + ...... ,\eqno{27}$$
 $$ B= B_0+\epsilon B_1 + \epsilon^2 B_2 + ....... , \eqno{28}$$
 $$ \phi = \epsilon \phi_1+ \epsilon^2 \phi_2 +..... \eqno {29}$$ 
Equation (25) gives us $ \phi_{1,r} = C tan 2r $ where C is a constant of integration. We choose $C={{1}\over{2}}$. The equation ( 22 )reduces at this order to 
$$ A_{,rr} - {8A \over {sin^2 (2r)}} =0 \eqno{30}$$
 which can be reduced to an equation of the hypergeometric type. A simple calculation shows that we have a special form of the hypergeometric equation, yielding 
$$ sin^{-1} (2r) _2F_1(-1/2,-1/2|-1/2|sin^2 (2r) )\eqno {31}$$
 for one solution, and 
$$ sin^2 (2r) _2F_1(1,1 | 5/2| sin^2 (2r)\eqno {32} $$
 for the other. These special forms of the hypergeometric function reduce to
 $$ A^1_1= cot (2r) ,\eqno{33}$$
 $$ A^2_1 = 3( 1-2r cot 2r) \eqno {34} $$
 respectively. For the former solution $ A^1_1$ we get two solutions 
$$ B_1^1 = cot 2r + tan 2r,\eqno {35} $$ 
$$ B_1^2 = -6r B_1^1 . \eqno {36}$$ 
By extending the perturbative analysis, for $A$ we get a differential equation, again of hypergeometric type , with an inhomogenous term, for $A$ which can be easily integrated. We are rather interested in the solution for $\phi$, to study its singularity structure. To second order in $\epsilon$ 
$$ \phi_{2,r} = {{12r-3 sin (4r)}\over { 2 cos^2 (2r)}}. \eqno {37} $$ 
In the expression found for $\phi$, there is no sign of singularity at $r=0$.

Our results may be interpreted better if we transform our original coordinates to 
$$ \overline {r} = tan (r) e^{B(r)} .\eqno {38}$$ 
Then our metric is transformed into 
$$ ds^2 = -(-M+ \overline {r}^2 )d \overline {t}^2 + {{d\overline {r}^2}\over {(-M+\overline {r}^2)}} +\overline{r}^2d\theta^2, \eqno {39}$$
 where 
$$ -M= e^{2(B-A)}[ sec^2(r) + 2 tan (r) B_{,r} + sin^2(r) B_{,r}^2 ] - tan^2 (r) e^{2B} \eqno{40} $$
 as given in the Pretorius-Choptuik reference .  Note that here $M$ is $r$ dependent.  In the original case $^{/9}$, it is also time dependent. Using our zeroth order solution for $A$, we find $\overline {r} = cot (2r) + O(\epsilon) .$ This transformation maps our original domain, $ 0 <\overline {r} <{{\pi}\over{4}} $ into  $ \infty >\overline {r}> 0 $. 

For the zeroth order solution $ -M= 1 $, a constant. We see that our solution corresponds to the AdS solution, known for this system. If we use the solution $B= log({{cos 2r} \over {2\alpha sin^2 r}})$, 
where $0<\alpha<1,$, we get the solutions corresponding to conical singularities $^{/38}$.

The BTZ  solution 
is obtained if we set $\alpha=-i$ which necessitates a reparametrization  using hyperbolic functions.  In the parametrization of the metric , eq.(21), hyperbolic cosine and hyperbolic tangent functions replace trigonometric cosine and trigonometric tangent respectively.  The presence of $i$ in the expression  in the expression for $B$  introduces the necessary minus sign for $-M$, eq.(39), while replacing $tan ( r ) $ by $tanh ( r ) $ in the metric, eq(21), retains the original signature.

 We, then, use our perturbative solutions in the presence of the scalar field. At first order in $\epsilon$, we find we have two solutions. If we take the solution set $A^1_1, B^1_1$, we find 
$$ -M=1+32\epsilon[ {{1}\over{\overline {r}}}- {\overline {r}}^3].\eqno {41}$$ 
This expression diverges both at $\overline {r}=0$ and $\overline {r}=\pi/4$ in an undesired fashion, so is discarded. For the latter solution set, $A^2_1, B^2_1$, 
$$ -M = 1 - 96\epsilon [ (1+\overline {r}^2)+ cot^{-1} (\overline {r})(-\overline {r}^3+{{1}\over{\overline {r}}})],\eqno {42} $$ 
an expression which diverges at the origin, but this time with the correct sign. At $\overline {r}=0 (r=\pi/4)$, $-M$ diverges to minus infinity. We excise the domain when $-M+\overline {r}^2$ equals zero. 

When the scalar field is calculated in terms of the new variable, we find 
$$\phi= 1/2 log \left({{(1+{\overline {r}}^2)^{1/2}}\over {\overline {r}}}\right). \eqno {43} $$ 
For $\overline {r} = 0 $, the scalar field $\phi$ is proportional to $log (\overline {r} ) $, as pointed out by Garfinkle $^ {/13}$
 and Burko $^{/39}$
. When $\overline {r} $ goes to infinity, $r=0$, $\phi$ goes to zero. The similar behaviour persists at second order in $\epsilon$, where 
$$ \phi_{2,r} = -3{{1}\over { \overline {r}}}- cot^{-1} (\overline {r}) ({{\overline {r}^2 +1} \over {\overline {r}^2}}).\eqno {44}$$ 
We see that the divergence is severer when $\overline {r}$ goes to zero. 

The fact that $-M$ goes to minus values as $\overline {r}$ approaches zero, signals the presence of a black hole around the origin. We can not tolerate $-M+\overline {r}^2 $ being null. We excise the space at the value of $\overline {r} $ where $-M+\overline {r}^2 $ equals zero. There is only one root of the equation 
$$ -M+{\overline {r}}^2=0, \eqno {45}$$
giving the approximate condition 
$$ \overline {r}> 24\pi \epsilon \left(1-(24)^2 {\pi}^2 {\epsilon}^2)\right)+O({\epsilon}^4) .\eqno {46}$$
 This will also prevent the curvature singularity which will occur at 
 $\overline {r}=0$. 

We have the scalar curvature made out of two parts , the finite part corresponding to the AdS solution and a singular part coming from the perturbative solution. 
$$ R = -6+{{\epsilon \pi}\over {\overline {r}^2}} [{{1}\over {1+\overline {r}^2}}],\eqno {47} $$ 
where $R$ is the curvature scalar given by Pretorius and Choptuik.

 To detect whether black-hole is formed or not, we use a second test given by Pretorius and Choptuik, by checking the condition for "trapped surfaces". In this reference, " trapped surface" is defined " to be surfaces where the expansion of the outgoing null curves normal to the surface is negative ". The condition for this to happen is given at the same place as 
$$ S= 1+ sin (r)cos (r) B_{,r} <0, \eqno{48  } $$
as applied to our case.
We study whether this constraint is satisfied for our solutions.
To zeroth order in $\epsilon$ this condition reads 
$$ 2S = -{{(1+(\overline{r}^2)^{1/2}}\over{\overline {r}}}. \eqno {49}$$
We see that "the surface is trapped" for $\overline{r} = 0 $ ($r=\pi/4$). 
For $r=0$ i.e. $\overline {r}$ approaching infinity, we get $S = - {{4}\over { \overline {r}^2}}$ which goes to zero from below. If we use the first order solution, however, the correction changes the situation in the first case, whereas it does not change the result in the second case.. For the first solution, we have
$$ S= 1+ sin(r) cos (r) B_{,r} =$$
$$ -[(1+(\overline{r})^2)^{1/2}\left({{1}\over {\overline{r}}}+\epsilon (1-{{1}\over {(\overline{r})^2}})\right) ] .\eqno {50}$$
For $\overline{r}/4 < \epsilon $ there is no "trapped surface. We can come all the way to the curvature singularity. This solution is discarded also for the above reason.
If we use the second solution, however, we get 
$$ S=
-(1+(\overline{r})^2)^{1/2}({{1}\over {\overline{r}}})$$
$$-3\epsilon\left[(1+(\overline{r})^2)^{-1/2}\left(\left(+\overline{r}+{{1}\over{\overline{r}}}\right)+cot^{-1}(\overline{r})\left({-\overline{r}}^2+{{1}\over{{\overline{r}}^2}}\right)\right)\right].\eqno {51}$$
As $\overline{r}$ goes to zero, the surface is trapped. The particle is not allowed to come close to the coordinate independent curvature singularity at the origin. 
\vskip .1truein

After we find the expressions given above for the scalar field and the metric, we use these approximate solutions, eq.s (35,36,43,44) and check whether they satisfy the set of exact equations, eq.s(22-25).The figures 1 to 4 of this reference $^{/37}$
show that although our approximate solutions do not satisfy the equations as $ \overline{r}$ goes to zero, the behaviour as $\overline{r}$ goes to infinity of these figures clearly show that asymptotically these solutions tend to be exact. We thus see that we can obtain approximate solutions to the equations of motion and the constraint equations, which approach exact solutions in the asymptotic region. Our tests, mainly the fact that $-M+\overline{r}^2$ goes through zero tells us that a black hole is formed as $\overline{r}$ goes to the origin (figure 5 of the reference). We excise the space at the point where $ -M+\overline{r}^2$ equals to zero. At this point note that our first set of approximate solutions, eq.s (33,35) do not give the correct asymptotic behaviour, so are discarded.

From these figures we see that our approximate solution is no longer reliable as $\overline{r}$ goes to zero. Since we excise our space in this region and the solution has the correct asymptotics as $\overline{r}$ goes to infinity, we think that the message our approximate solution conveys, i.e. the presence of a black hole at the origin, is correct.
\vfill\eject
\noindent
Final Remarks
\vskip.05truein
Here I tried to give a taste of a subject Matt Choptuik started a decade ago.  there is a large literature built around this work nowadays.  The original Choptuik paper $^{5}$ has over three hundred citations in ten years, which may not be a lot compared to a paper on a particle physics experiment, but is considerable, when compared to the citations of one of the best known reviews in general relativity, namely the Eguchi, Gilkey, Hanson paper $^{/40}$(670 citations in twenty three years).  This topic is considered one of the areas where most of the work  numerical relativity  is done.
\vskip.1truein
\noindent
{\bf{Acknowledgement}}:
I thank Prof. Aliev for giving me this opportunity to talk in this "Research Semester on Gravitation and Cosmology" to such a distinguished audience.  I also thank TUBITAK, the Scientific and Technical Council of Turkey, and TUBA, the Academy of Sciences of Turkey for partial support.

\vskip 1 truein
\noindent
{{\bf{REFERENCES}}}
\item {1.} 
 Carsten Gundlach, Physics Reports, {\bf{376}}(2003)339; 
, gr-qc/0210101 

\item {2.}
 Carsten Gundlach, 
 Living Reviews Rel.{\bf{2}}(1999)4;
, gr-qc/0001046 

\item {3.}
 Carsten Gundlach , 
, Adv.Theor.Math.Phys.{\bf{2}}(1998)1; 
, gr-qc/9712084 

\item {4.} Carsten Gundlach, Physical ReviewD {\bf{55}} (1997)695; 

\item {5.} Matthew W. Choptuik, Physical Review Letters,{\bf {40}} (1993) 9;

\item {6.}  A.M.Abrahams and C.R.Evans ,Physical Review Letters {\bf{70}} (1993) 2980;

\item {7.}  C.R.Evans and J.S.Coleman ,Physical Review Letters {\bf{72}} (1994)1782;

\item {8.}   Jason F. Ventrella , Matthew W. Choptuik, 
 gr-qc/0304007 ;

\item {9.} Frans Pretorius and Matthew W. Choptuik, Physical Review {\bf{D 62}} (2000) 124012;

\item {10.} Viqar Husain and Michel Olivier, Classical and Quantum Gravity,{\bf{ 18}} (2001) L1;

\item {11.} Andrei V. Frolov, Classical and Quantum Gravity, {\bf{16}} (1999) 407;

\item {12.} G\' erard Cl\' ement and Alessandro Fabbri, gr-qc/0101073;

\item {13.} D. Garfinkle, Physical Review {\bf{D 63}} (2001) 044007;

\item {14.}   D.Birmingham, Phys. Rev. D {\bf{64}} (2001) 064024, hep-th/0101194;

\item {15.} G.T.Horowitz and D.L.Welch, Physical Review Letters, {\bf{71}} (1993) 328;

\item {16.} N.Kaloper, Physical Review D {\bf{48}} (1993) 2598;

\item {17.} Abbas Ali, Alok Kumar,  
 Modern Physics Letters A {\bf{8}} (1993), hep-th/9303032;

\item {18.}   J.Maldacena, Adv. Theo. Math. Phys. {\bf{2}} (1998) 231;

\item{19.} D.Friedan, Z.Qui and S. Shenker, Physical Review Letters {\bf{52}} (1984) 1575;

\item {20.} A.A.Belavin, A.M.Polyakov and A.B. Zamolodchikov, Nuclear Physics,{\bf{ B241}} (1984) 333;

\item {21.} Gerald t'Hooft, Physical Review {\bf{D 14}} (1976) 3432;

\item {22.} N.Seiberg and E. Witten, Nuclear Physics {\bf{B426}} (1994) 19;

\item {23.} Adel Bilal, {\it{Duality in N=2 SUSY SU(2) Yang-Mills Theory}}, hep-th/9601007;

\item {24.}  Matthew W. Choptuik (, Eric W. Hirschmann , Steven L. Liebling, Frans Pretorius, 
 gr-qc/0305003; 

\item {25.} D.Christodolou, Communications in Mathematical Physics {\bf{105}} (1986) 337; 

\item {26.} D.Christodolou, Communications in Pure and Applied Mathematics {\bf{46}} (1993) 1131);

\item {27.} D.Christodolou, Communications in  Pure and  Applied Mathematics {\bf{44}} (1991) 339) ;

\item {28.}  J.M. Martin-Garcia and C. Gundlach, gr-qc/0304070;

\item {29.} M.Banados, C.Teitelboim and Jorge Zanelli, Physical Review Letters, {\bf{69}} (1992) 1849; also, M.Banados, M.Henneaux, C.Teitelboim and J.Zanelli, Physical Review, {\bf{D48}} (1993) 1506;

\item{30.} D.Garfinkle and C.Gundlach, Physical Review, D {\bf{66}} (2002) 044015;

\item {31.}  E.W. Hirschmann, A. Wang and Y.Wu, gr-qc/0207121;

\item {32.} D.Birmingham and S.Sen, Physical Review Letters {\bf{84}} (2000) 1074;

\item {33.} Y.Peleg and A.Steif, Physical Review D {\bf{51}} (1995) 3992 ;

\item {34.} D.Birmingham, Physical Review D {\bf{64}} (2001) 064024;

\item {35.}  D. Birmingham, Ivo Sachs and Siddartha Sen, hep-th/0102155; 

\item {36.} Cristian Martinez, C.Teitelboim and Jorge Zanelli, Physical Review {\bf{D61}} (2000) 1849104013.

\item {37.} T.Birkandan and M.Horta\c csu, Gen.Rel.Grav. {\bf{35}}(2003)457,gr-qc/0104096);

\item {38.} S. Carlip, Classical and Quantum Gravity, {\bf {12}} (1995) 2853;

\item {39.} Lior M. Burko, Physical Review {\bf{D 62}} (2000) 127503;

\item {40.}  Tohru Eguchi , Peter B. Gilkey , Andrew J. Hanson,
 Physics Reports, {\bf{66}}(1980)213. 

\end

\end